\documentclass[12pt]{iopart}

\usepackage{graphicx}
\usepackage{fancyhdr}
\usepackage{amsfonts,amssymb,color}

\definecolor{red}{rgb}{1,0,0}

\begin{document}

\title[]{Anomaly-free scalar perturbations with holonomy corrections  
in loop quantum cosmology}

\author{Thomas Cailleteau$^1$, Jakub Mielczarek$^2$, Aurelien Barrau$^1$ and Julien Grain$^3$}

\address{$^1$ Laboratoire de Physique Subatomique et de Cosmologie, UJF, CNRS/IN2P3, INPG\\
53, av. des Martyrs, 38026 Grenoble cedex, France}
\address{$^2$ Astronomical Observatory, Jagiellonian University, \\30-244
Krak\'ow, Orla 171, Poland}
\address{$^2$ Theoretical Physics Department, National Centre for Nuclear Research, \\ Ho{\.z}a 69, 00-681 Warsaw, Poland}
\address{$^3$ Institut d'Astrophysique Spatiale, Universit\'e Paris-Sud 11, CNRS \\ B\^atiments 120-121, 
91405 Orsay Cedex, France}

\begin{abstract}
Holonomy corrections to scalar perturbations are investigated in the loop quantum cosmology  framework.
Due to the effective approach, modifications of the algebra of constraints generically lead to anomalies.
In order to remove those anomalies, counter-terms are introduced. We find a way to explicitly fulfill the
conditions for anomaly freedom and we give explicit expressions for the counter-terms. Surprisingly, the $\bar{\mu}$-scheme naturally arises in this 
procedure. The gauge invariant variables are found and equations of motion for the anomaly-free scalar perturbations 
are derived.   Finally, some cosmological consequences are discussed qualitatively. 
\end{abstract}

\maketitle

\section{Introduction}

Loop Quantum Gravity (LQG) is a tentative non-perturbative and background-independent 
quantization of General Relativity (GR) \cite{lqg_review}. Interestingly, it has now been demonstrated that different
approaches, based on canonical quantization of GR, on covariant quantization of GR and on formal quantization 
of geometry lead to the very same LQG theory. Although this is rather convincing, a direct experimental probe is 
still missing. One can easily argue that cosmology is the most promising approach to search for observational features 
of LQG or, more specifically, to its symmetry-reduces version, Loop Quantum Cosmology (LQC) \cite{lqc_review}.

Many efforts have been devoted to the search of possible footprints of LQC in cosmological tensor modes (see \cite{tensor}).
At the theoretical level, the situation is easier in this case as the algebra of constraints is automatically anomaly-free. 
But, as far as observations are concerned, scalar modes are far more important. They have already been observed in 
great details by WMAP \cite{wmap} and are currently even better observed by the Planck mission. The question of a possible 
modification of the primordial scalar power spectrum (and of the corresponding TT $C_l$ spectrum) in LQC is therefore essential 
in this framework.

Gravity is described by a set of constraints. However, for the (effective) theory to be consistent, it is mandatory that the
evolution generated by the constraints remains compatible with the constraints themselves. This is always true if their mutual Poisson
brackets vanish when evaluated in fields fulfilling the constraints, {\it i.e.} if they form a first class algebra. This
means that the evolution and the gauge transformations are associated with vector fields that are tangent to the manifold of
null constraints. This obviously holds at the classical level. However, when quantum modifications are added, the anomaly
freedom is not anymore automatically ensured. Possible quantum corrections must be restricted to those which close the
algebra. This means that, for consistency reasons, the Poisson brackets between any two constraints must be proportional to
one constraint of the algebra. This article is devoted to the search for such an algebra for scalar perturbations. 

Our approach will follow the one developed by Bojowald {\it et al.} in \cite{Bojowald:2008gz}. There are two main quantum
corrections expected from LQC: inverse volume terms, basically arising for inverse powers of the densitized triad, which when 
quantized become an operator with zero in its discrete spectrum thus lacking a direct inverse, and holonomy corrections coming
from the fact that loop quantization is based on holonomies, rather than direct connection components. In \cite{Bojowald:2008gz} 
the authors focused exclusively on inverse volume corrections. Here, we extend with work to the holonomy corrections. 
Scalar perturbations with holonomy corrections have been studied in \cite{Wu:2010wj}. However, the issue of anomaly freedom 
was not really addressed. Recently, a new  possible way of introducing holonomy corrections  to the scalar perturbations  was 
proposed in \cite{WilsonEwing:2011es}.  Although it was interestingly shown that the formulation is anomaly-free, the approach is based on 
the choice of the longitudinal gauge and the extension of the method to the gauge-invariant case is not straightforward. In contrast, 
the approach developed in our paper does not rely on any particular choice of gauge and the gauge-invariant 
cosmological perturbations are easily constructed. 
    
The theory of anomaly-free scalar perturbations  developed in this paper is  obtained on a flat FRW background, such
that the line element is given by:
\begin{equation}
ds^2 = a^2\left[ -(1+2\phi)d\eta^2+2\partial_a B d\eta dx^a+ ((1-2\psi)\delta_{ab}+2\partial_a \partial_b E)dx^adx^b  \right],
\label{lineelement}
\end{equation}
where $\phi$, $\psi$, $E$ and $B$ are scalar perturbation functions. The matter content 
is assumed to be a scalar field. This will allow us to investigate the generation of scalar 
perturbations during the phase of cosmic inflation while taking into account the quantum gravity effects. 

Our analysis of the scalar perturbations is performed in the Hamiltonian framework developed in
\cite{Bojowald:2008gz,Bojowald:2008jv}.  
As it was shown there, the background variables are $(\bar{k},\bar{p},\bar{\varphi},\bar{\pi})$, while the perturbed 
variables are $(\delta K^i_a,\delta E^a_i, \delta \varphi, \delta \pi)$. The Poisson bracket for the system can be decomposed 
as follows:
\begin{equation}
\{ \cdot , \cdot \} =  \{ \cdot, \cdot \}_{\bar{k},\bar{p}} + \{\cdot, \cdot \}_{\delta K, \delta E} 
+ \{\cdot,\cdot \}_{\bar{\varphi},\bar{\pi}} + \{\cdot,\cdot \}_{\delta \varphi, \delta \pi}
\label{Poisson}
\end{equation}
where 
\begin{eqnarray}
\{ \cdot, \cdot\}_{\bar{k},\bar{p}} &:=&\frac{\kappa}{3 V_0} \left[ \frac{\partial  \cdot}{\partial \bar{k}} \frac{\partial  \cdot}{\partial \bar{p}} 
-\frac{\partial  \cdot}{\partial \bar{p}} \frac{\partial  \cdot}{\partial \bar{k}}\right], \\
\{\cdot, \cdot \}_{\delta K, \delta E} &:=&\kappa \int_{\Sigma} d^3x \left[ \frac{\delta  \cdot}{\delta \delta K^i_a} \frac{\delta  \cdot}{\delta \delta E^a_i} 
-\frac{\delta  \cdot}{\delta \delta E^a_i} \frac{\delta  \cdot}{\delta \delta K^i_a}\right], \\
\{\cdot, \cdot \}_{\bar{\varphi},\bar{\pi}} &:=&\frac{1}{V_0} \left[ \frac{\partial  \cdot}{\partial \bar{\varphi}} \frac{\partial  \cdot}{\partial \bar{\pi}} 
-\frac{\partial  \cdot}{\partial \bar{\pi}} \frac{\partial  \cdot}{\partial \bar{\varphi}}\right], \\
\{\cdot, \cdot \}_{\delta \varphi, \delta \pi} &:=&\int_{\Sigma} d^3x  \left[ \frac{\delta  \cdot}{\delta \delta \varphi} \frac{\delta  \cdot}{\delta \delta \pi} 
-\frac{\delta  \cdot}{\delta \delta \pi} \frac{\delta  \cdot}{\delta \delta \varphi}\right]. 
\end{eqnarray}
Here, $V_0$ is the volume of the fiducial cell and $\kappa=8\pi G$. 

The holonomy corrections are introduced by the replacement $\bar{k} \rightarrow \mathbb{K}[n]$ in the classical Hamiltonian. We follow the notation introduced in \cite{VectorJTAJ}, where
\begin{equation}
\mathbb{K}[n] := \left\{  
\begin{tabular}{ccc} 
$\frac{\sin(n\bar{\mu} \gamma \bar{k})}{n\bar{\mu}\gamma}$  &  for &$n \in \mathbb{Z}/\{0\}$,  \\ 
& & \\
 $\bar{k}$  &  for & $n=0$,
\end{tabular}
\right.
\end{equation}
for the correction function. In cases where $\bar{k}$ appears quadratically, the integer 
$n$ is fixed to  two (See \cite{VectorJTAJ}). In the other cases, the integers remain to be fixed 
from the requirement of anomaly freedom. The coefficient $\gamma$ is the Barbero-Immirzi parameter 
and $\bar{\mu} \propto \bar{p}^{\beta}$  where $-1/2  \leq  \beta  \leq 0 $.  In what follows, the relation 
\begin{equation}
\bar{p} \frac{\partial}{\partial \bar{p}}\mathbb{K}[n]  = [\bar{k} \cos(n \bar{\mu}\gamma \bar{k})-\mathbb{K}[n] ]\beta  
\end{equation}
will be useful.   

The organization of the paper is the following. In  Sec. \ref{Sperthol}, the holonomy-corrected gravitational  
Hamiltonian constraint is defined. We calculate the Poisson bracket of the Hamiltonian constraint with itself and 
with the gravitational diffeomorphism constraint. In Sec. \ref{matter}, scalar matter is introduced. 
The Poisson brackets between the total constraints for the system under consideration are calculated. 
In Sec. \ref{anomaly}, the conditions for anomaly freedom are solved and the expressions for the 
counter-terms are derived. Based on this, in Sec. \ref{equations}, equations of motion for the scalar 
perturbations are derived. The system of equations is then investigated in the case of the longitudinal 
gauge. Finally, gauge-invariant variables are found and the equations for the corresponding Mukhanov
variables are derived.  In Sec. \ref{summary}, we summarize our results and  draw out some conclusions. 

\section{Scalar perturbations with holonomy corrections} \label{Sperthol}

The holonomy-modified Hamiltonian constraint can be written as: 
\begin{equation}
H^Q_G[N] = \frac{1}{2 \kappa} \int_{\Sigma} d^3x \left[ \bar{N}(\mathcal{H}^{(0)}_G+\mathcal{H}^{(2)}_G)+\delta N \mathcal{H}^{(1)}_G\right],
\label{HolModHam}
\end{equation}
where 
\begin{eqnarray}
\mathcal{H}^{(0)}_G = -6 \sqrt{\bar{p}} (\mathbb{K}[1])^2, \nonumber \\
\mathcal{H}^{(1)}_G  = -4\sqrt{\bar{p}} \left(  \mathbb{K}[s_1] +\alpha_1\right)
\delta^c_j \delta K^j_c-\frac{1}{\sqrt{\bar{p}}} 
\left(  \mathbb{K}[1]^2+\alpha_2  \right) \delta^j_c\delta E^c_j \nonumber \\
+\frac{2}{\sqrt{\bar{p}}} (1+\alpha_3) \partial_c \partial^j \delta E^c_j,  \nonumber  \\
\mathcal{H}^{(2)}_G  = \sqrt{\bar{p}}(1+\alpha_4) \delta K^j_c \delta K^k_d \delta^c_k \delta^d_j 
-\sqrt{\bar{p}} (1+\alpha_5)(\delta K^j_c \delta^c_j)^2 \nonumber \\
-\frac{2}{\sqrt{\bar{p}}} \left(  \mathbb{K}[s_2]+\alpha_6 \right) \delta E^c_j \delta K^j_c   
-\frac{1}{2\bar{p}^{3/2}} \left( \mathbb{K}[1]^2+\alpha_7\right)\delta E^c_j \delta E^d_k \delta^k_c \delta^j_d \nonumber \\
+\frac{1}{4\bar{p}^{3/2}}\left( \mathbb{K}[1]^2+\alpha_8\right) (\delta E^c_j \delta^j_c)^2 
- \frac{1}{2\bar{p}^{3/2}}(1+\alpha_9) \delta^{jk} (\partial_c \delta E^c_j)(\partial_d \delta E^d_k). \nonumber 
\end{eqnarray}
The standard holonomy corrections are parametrized by two integers $s_1$ and $s_2$. 
The $\alpha_i$ are counter-terms, which are introduced to remove anomalies. 
Those factors are defined so that they vanish in the classical limit $(\bar{\mu} \rightarrow 0)$.
The counter-terms could be, in general,  functions of all the canonical variables. 
We however assume here that they are functions of the gravitational background 
variables only. 
  
In our approach, the diffeomorphism constraint holds the classical form
\begin{equation}
D_G[N^a] = \frac{1}{\kappa} \int_{\Sigma} d^3x \delta N^c \left[ \bar{p} \partial_c(\delta^d_k \delta K^k_d )
-\bar{p}(\partial_k \delta K^k_c)
-\bar{k} \delta^k_c (\partial_d \delta E^d_k )\right].
\label{diffconstr}
\end{equation}
In general, the diffeomorphism constraint could also be holonomy corrected. 
this possibility was studied, {\it e.g.}, in \cite{Wu:2010wj}. However, in LQG the 
diffeomorphism constraint is satisfied at the classical level. Therefore, if LQC is to be considered as 
a specific model of LQG, the  diffeomorphism constraint should naturally hold its classical form. Because of this, 
in this paper, the  diffeomorphism  constraint is not modified by the holonomies.   
It is worth stressing, that the classicality of the diffeomorphism constraint 
is also imposed by the requirement of anomaly cancelation. Namely, if one replaces $ \bar{k} \rightarrow \mathbb{K}[n]$
in (\ref{diffconstr}), the condition $n=0$ would anyway be required by the introduction of scalar matter. 
In fact, the same condition was obtained for vector modes with holonomy 
corrections  \cite{VectorJTAJ}. 

Let us now calculate the possible Poisson brackets for the constraints $H^Q_G[N]$ and $D_G[N^a]$. 

\subsection{The $\left\{H^Q_G, D_G \right\}$ bracket}

Using the definition of the Poisson bracket (\ref{Poisson}), we derive:
\begin{eqnarray}
 \left\{H_{G}^Q[N],D_{G}[N^a] \right\} = - H_{G}^Q\left[ \delta N^a \partial_a \delta N \right] +\mathcal{B}\ D_{G}[N^a]    \nonumber  \\ 
+ \frac{\sqrt{\bar{p}}}{\kappa}\int_{\Sigma}  d^3x \delta N^a (\partial_a \delta N) \mathcal{A}_1 
+ \frac{\bar{N}\sqrt{\bar{p}}\bar{k} }{\kappa}\int_{\Sigma}  d^3x \delta N^a (\partial_i \delta K^i_a )  \mathcal{A}_2 \nonumber  \\ 
+\frac{\bar{N}}{\kappa \sqrt{\bar{p}}}\int_{\Sigma}  d^3x \delta N^i (\partial_a \delta E^a_i)  \mathcal{A}_3 
+\frac{\bar{N}}{2 \kappa \sqrt{\bar{p}}}\int_{\Sigma}  d^3x  (\partial_a \delta N^a) (\delta E^b_i \delta^i_b)  \mathcal{A}_4, 
\label{PoissHGDG}
\end{eqnarray}
where 
\begin{equation}
\mathcal{B}  = \frac{\bar{N}}{\sqrt{\bar{p}}} \left[ -2\mathbb{K}[2]+\bar{k}(1+\alpha_5)+\mathbb{K}[s_2]+\alpha_6  \right],
\label{Bfunction}
\end{equation}
and 
\begin{eqnarray}
\mathcal{A}_1 =  2\bar{k}(\mathbb{K}[s_1]+\alpha_1)+\alpha_2-2\mathbb{K}[1]^2,  \\  
\mathcal{A}_2 = \alpha_5-\alpha_4,   \\  
\mathcal{A}_3 = -\mathbb{K}[1]^2- \bar{p} \frac{\partial}{\partial \bar{p}}\mathbb{K}[1]^2-\frac{1}{2}\alpha_7 \nonumber \\
+\bar{k}(-2\mathbb{K}[2]+\bar{k}(1+\alpha_5)+2\mathbb{K}[s_2]+2\alpha_6),     \\  
\mathcal{A}_4 = \alpha_8-\alpha_7.  
\end{eqnarray}

The functions $\mathcal{A}_1,\dots,  \mathcal{A}_4$  are the first anomalies coming from the effective 
nature of the Hamiltonian constraint. Later, we will set them to zero so as to fulfill the requirement of anomaly 
freedom. This will lead to constraints on the form of the counter-terms. 

Beside the anomalies, the $\left\{H^Q_G, D_G \right\}$ bracket contains the $- H_{G}^Q\left[ \delta N^a \partial_a \delta N \right]$
term, which is expected classically.  There is also an additional contribution from the diffeomorphism constraint $\mathcal{B}\ D_{G}[N^a]$.
This term is absent in the classical theory. This is however consistent as, for $\bar{\mu} \rightarrow 0$, the $\mathcal{B}$ function tends to zero.    

\subsection{The $\left\{H^Q_G, H^Q_G \right\}$ bracket}

The next bracket is:
\begin{eqnarray}
 \left\{H_{G}^Q[N_1],H_{G}^Q[N_2] \right\} = 
 (1+\alpha_3)(1+\alpha_5)D_G \left[ \frac{\bar{N}}{\bar{p}} \partial^a(\delta N_2 -\delta N_1)  \right] \nonumber \\
+\frac{\bar{N}}{\kappa} \int_{\Sigma} d^3x \partial^a(\delta N_2 -\delta N_1)(\partial_i \delta K^i_a)(1+\alpha_3)\mathcal{A}_5 \nonumber \\
+\frac{\bar{N}}{\kappa \bar{p}} \int_{\Sigma} d^3x(\delta N_2 -\delta N_1)(\partial^i\partial_a \delta E^a_i) \mathcal{A}_6 \nonumber \\
+\frac{\bar{N}}{\kappa} \int_{\Sigma} d^3x(\delta N_2 -\delta N_1)(\delta^a_i  \delta K^i_a) \mathcal{A}_7 \nonumber \\
+\frac{\bar{N}}{\kappa \bar{p}} \int_{\Sigma} d^3x (\delta N_2 -\delta N_1)(\delta^i_a \delta E^a_i) \mathcal{A}_8, 
\end{eqnarray}
where 
\begin{eqnarray}
\mathcal{A}_5 =  \alpha_5 - \alpha_4,   \\  
\mathcal{A}_6 =  (1+\alpha_9)(\mathbb{K}[s_1]+\alpha_1)-(1+\alpha_3)(\mathbb{K}[s_2]+\alpha_6)+\mathbb{K}[2](1+\alpha_3) \nonumber  \\
-2\mathbb{K}[2]\bar{p}\frac{\partial \alpha_3}{\partial \bar{p}}
+ \frac{1}{2}\left(  \mathbb{K}[1]^2+2\bar{p}\frac{\partial}{\partial \bar{p}}\mathbb{K}[1]^2  \right)\frac{\partial \alpha_3}{\partial \bar{k}}
-\bar{k}(1+\alpha_3)(1+\alpha_5),   \\
\mathcal{A}_7 = 4\mathbb{K}[2] \bar{p} \frac{\partial }{\partial \bar{p}} (\mathbb{K}[s_1]+\alpha_1)
-\left(  \mathbb{K}[1]^2+2\bar{p}\frac{\partial}{\partial \bar{p}}\mathbb{K}[1]^2  \right) \frac{\partial }{\partial \bar{k}} (\mathbb{K}[s_1]+\alpha_1)
\nonumber   \\  
+\left(1+ \frac{3}{2}\alpha_5-\frac{1}{2}\alpha_4\right)(\mathbb{K}[1]^2+\alpha_2)
-2(\mathbb{K}[s_2]+\alpha_6)(\mathbb{K}[s_1]+\alpha_1) \nonumber \\
+2\mathbb{K}[2](\mathbb{K}[s_1]+\alpha_1), \label{A7} \\
\mathcal{A}_8 = \frac{1}{2}(\mathbb{K}[s_2]+\alpha_6)(\mathbb{K}[1]^2+\alpha_2)
-(\mathbb{K}[s_1]+\alpha_1)(\mathbb{K}[1]^2+\alpha_7) \nonumber \\
+\frac{3}{2}(\mathbb{K}[s_1]+\alpha_1)(\mathbb{K}[1]^2+\alpha_8) 
-\frac{1}{2} \mathbb{K}[2](\mathbb{K}[1]^2+\alpha_2) \nonumber \\
+\mathbb{K}[2]\bar{p} \frac{\partial}{\partial \bar{p}}(\mathbb{K}[1]^2+\alpha_2)
-\frac{1}{4}\left( \mathbb{K}[1]^2+2\bar{p} \frac{\partial}{\partial \bar{p}} \mathbb{K}[1]^2\right)
\frac{\partial }{\partial \bar{k}}(\mathbb{K}[1]^2+\alpha_2). \label{A8}
\end{eqnarray}

The $\mathcal{A}_5,\dots,  \mathcal{A}_8$ are the next four anomalies.  Moreover, the diffeomorphism
constraint is multiplied by the factor $(1+\alpha_3)(1+\alpha_5)$.
  
\subsection{The $\left\{D_G, D_G \right\}$ bracket}

The Poisson bracket between the diffeomorphism constraints is:
\begin{eqnarray}
 \left\{D_{G}[N^a_1],D_{G}[N^a_2] \right\} = 0.
\end{eqnarray}

\section{Scalar matter} \label{matter}

In this section, we introduce scalar matter.
The scalar matter diffeomorphism constraint is 
\begin{equation} 
D_M[N^a] = \int_{\Sigma} \delta N^a \bar{\pi} (\partial_a \delta \varphi).
\end{equation}
The scalar matter Hamiltonian can be expressed as:
\begin{equation}
H^Q_M[N]=H_M[\bar{N}] +H_M[\delta N],\nonumber
\end{equation}
where 
\begin{eqnarray}
H_M[\bar{N}] &=& \int_{\Sigma} d^3 x \bar{N} \left[ \left(\mathcal{H}^{(0)}_{\pi}+\mathcal{H}^{(0)}_{\varphi}\right)
+\left(\mathcal{H}^{(2)}_{\pi}+\mathcal{H}^{(2)}_{\nabla}+\mathcal{H}^{(2)}_{\varphi}  \right)  \right],  \label{HMb} \\
H_M[\delta N] &=&  \int_{\Sigma} d^3 \delta N \left[  \mathcal{H}^{(1)}_{\pi}+\mathcal{H}^{(1)}_{\varphi} \right]. \label{HMd} \
\end{eqnarray}
The factors in equations (\ref{HMb}) and  (\ref{HMd}) are 
\begin{eqnarray}
\mathcal{H}^{(0)}_{\pi} &=& \frac{\bar{\pi}^2}{2\bar{p}^{3/2}}, \nonumber \\
\mathcal{H}^{(0)}_{\varphi} &=& \bar{p}^{3/2} V(\bar{\varphi}), \nonumber \\
\mathcal{H}^{(1)}_{\pi} &=&  
\frac{\bar{\pi} \delta \pi}{\bar{p}^{3/2}}-\frac{\bar{\pi}^2}{2\bar{p}^{3/2}} \frac{\delta^j_c \delta E^c_j}{2\bar{p}}, \nonumber  \\
\mathcal{H}^{(1)}_{\varphi} &=& 
\bar{p}^{3/2} \left[ V_{,\varphi}(\bar{\varphi}) \delta \varphi +V(\bar{\varphi}) \frac{\delta^j_c \delta E^c_j}{2\bar{p}} \right], \nonumber \\
\mathcal{H}^{(2)}_{\pi} &=& \frac{1}{2} \frac{\delta \pi^2}{\bar{p}^{3/2}}-\frac{\bar{\pi} \delta \pi}{\bar{p}^{3/2}} \frac{\delta^j_c \delta E^c_j}{2\bar{p}}
+\frac{1}{2} \frac{\bar{\pi}^2}{\bar{p}^{3/2}} \left[  \frac{(\delta^j_c \delta E^c_j )^2}{8\bar{p}^2}
+\frac{\delta^k_c \delta^j_d \delta E^c_j \delta E^d_k}{4\bar{p}^2}  \right],  \\
\mathcal{H}^{(2)}_{\nabla} &=& \frac{1}{2} \sqrt{\bar{p}}(1+\alpha_{10}) 
\delta^{ab} \partial_a \delta \varphi  \partial_b \delta \varphi,\nonumber  \\
\mathcal{H}^{(2)}_{\varphi} &=& \frac{1}{2}  \bar{p}^{3/2} V_{,\varphi\varphi}(\bar{\varphi}) \delta \varphi^2 
+\bar{p}^{3/2} V_{,\varphi}(\bar{\varphi}) \delta \varphi  \frac{\delta^j_c \delta E^c_j}{2\bar{p}}  \\
&+&\bar{p}^{3/2} V(\bar{\varphi}) \left[  \frac{(\delta^j_c \delta E^c_j )^2}{8\bar{p}^2}
-\frac{\delta^k_c \delta^j_d \delta E^c_j \delta E^d_k}{4\bar{p}^2}  \right]. 
\end{eqnarray}
Here, we have introduced the counter-term $\alpha_{10}$  in the factor $\mathcal{H}^{(2)}_{\nabla}$. 
Thanks to this, the Poisson bracket between two matter Hamiltonians takes the following form:
\begin{equation}
\left\{H_M^Q[N_1],H_M^Q[N_2]\right\} = (1+\alpha_{10}) D_M \left[ \frac{\bar{N}}{\bar{p}}\partial^a(\delta N_2-\delta N_1)  \right].
\end{equation}
As  will be explained later, the appearance of the front-factor $(1+\alpha_{10}) $ will allow us to close 
the algebra of total constraints. In principle, other prefactors could have been expected, however they do not help removing anomalies. 

\subsection{Total constraints}

The total Hamiltonian and diffeomorphism constraints are the following:
\begin{eqnarray}
H_{tot}[N] &=& H^Q_G[N]+H^Q_M[N],   \\
D_{tot}[N^a] &=& D_G[N^a]+D_M[N^a]. 
\end{eqnarray}
The Poisson bracket between two total  diffeomorphism constraints is vanishing:
\begin{equation}
\left\{D_{tot}[N^a_1],D_{tot}[N^a_2]\right\}  = 0.
\end{equation}
The bracket between  the total Hamiltonian and diffeomorphism constraints can 
be decomposed as follows:
\begin{eqnarray}
\left\{ H_{tot}[N],D_{tot}[N^a] \right\}  &=& \left\{ H_{M}^Q[N],D_{tot}[N^a] \right\} +\left\{ H_{G}^Q[N],D_{G}[N^a] \right\} \nonumber \\
&+&\left\{ H_{G}^Q[N],D_{M}[N^a] \right\}. \label{decompHtotDtot}
\end{eqnarray}
The first bracket in the sum (\ref{decompHtotDtot}) is given by 
\begin{equation}
\left\{ H^Q_{M}[N],D_{tot}[N^a] \right\} = - H^Q_M[\delta N^a \partial_a \delta N].
\end{equation}
The second contribution to Eq. (\ref{decompHtotDtot}) is given by  (\ref{PoissHGDG}),
while the last contributions is vanishing:
\begin{equation}
\left\{ H_{G}^Q[N],D_{M}[N^a] \right\} = 0.
\end{equation}
The Poisson bracket between the two total Hamiltonian constraints can be decomposed in the following way:
\begin{eqnarray}
\left\{ H_{tot}[N_1],H_{tot}[N_2] \right\}  &=& \left\{ H_G^Q[N_1],H_G^Q[N_2] \right\} +\left\{ H_M[N_1],H_M[N_2] \right\} \nonumber \\
     &+& \left[  \left\{H_G^Q[N_1],H_M[N_2]\right\} - (N_1 \leftrightarrow N_2)  \right].
\end{eqnarray}
The contribution from the last brackets can be expressed as
\begin{eqnarray}
\left\{H_G^Q[N_1],H_M[N_2]\right\} - (N_1 \leftrightarrow N_2)  =  \nonumber \\
=\frac{1}{2} \int_{\Sigma}d^3x \bar{N} (\delta N_2-\delta N_1)\left( \frac{\bar{\pi}^2}{2\bar{p}^3}-V(\bar{\varphi}) \right)
(\partial_c\partial^j \delta E^c_j) \mathcal{A}_9 \nonumber \\
+ 3 \int_{\Sigma}d^3x \bar{N} (\delta N_2-\delta N_1)\left( \frac{\bar{\pi}\delta \pi}{\bar{p}^2}-\bar{p}V_{\varphi}(\bar{\varphi}) \delta \varphi \right)
\mathcal{A}_{10} \nonumber \\
+\int_{\Sigma}d^3x \bar{N} (\delta N_2-\delta N_1)(\delta^c_j \delta K^c_j) \left( \frac{\bar{\pi}^2}{2\bar{p}^3}-V(\bar{\varphi}) \right) \bar{p}
 \mathcal{A}_{11} \nonumber \\
+\frac{1}{2} \int_{\Sigma}d^3x \bar{N} (\delta N_2-\delta N_1)(\delta^j_c \delta E^c_j) \left(\frac{\bar{\pi}^2}{2\bar{p}^3}\right) 
\mathcal{A}_{12} \nonumber \\
+\frac{1}{2} \int_{\Sigma}d^3x \bar{N} (\delta N_2-\delta N_1)(\delta^j_c \delta E^c_j) V(\bar{\varphi}) 
\mathcal{A}_{13},
\end{eqnarray}
where
\begin{eqnarray}
\mathcal{A}_9 = \frac{\partial \alpha_3}{\partial \bar{k}},  \\
\mathcal{A}_{10} = \mathbb{K}[2]-\mathbb{K}[s_1]-\alpha_1, \\
\mathcal{A}_{11} = -\frac{\partial}{\partial \bar{k}}(\mathbb{K}[s_1]+\alpha_1)+\frac{3}{2}(1+\alpha_5)-\frac{1}{2}(1+\alpha_4), \\
\mathcal{A}_{12} = -\frac{1}{2}\frac{\partial}{\partial \bar{k}}(\mathbb{K}[1]^2+\alpha_2)+5(\mathbb{K}[s_1]+\alpha_1)
- 5\mathbb{K}[2]+\mathbb{K}[s_2] +\alpha_6, \label{A12} \\
\mathcal{A}_{13} = \frac{1}{2}\frac{\partial}{\partial \bar{k}}(\mathbb{K}[1]^2+\alpha_2)+\mathbb{K}[s_1]+\alpha_1 
-\mathbb{K}[2]-\mathbb{K}[s_2]-\alpha_6.
\end{eqnarray}
The functions $\mathcal{A}_9,\dots,  \mathcal{A}_{13}$ are the last five anomalies. 

\section{Anomaly freedom} \label{anomaly}

The requirement of anomaly freedom is equivalent to the
conditions $\mathcal{A}_i=0$ for $i=1,\dots, 13$. 

Let us start form the condition $\mathcal{A}_9=0$. Since $\alpha_3$ cannot be 
a constant, this condition implies $\alpha_3=0$.  The condition
$\mathcal{A}_{10}=0$ gives $\alpha_1 =  \mathbb{K}[2]-\mathbb{K}[s_1]$.
Using this, the condition $\mathcal{A}_{1}=0$, can be written 
as $\alpha_2 =  2\mathbb{K}[1]^2 -2\bar{k}\mathbb{K}[2]$. The conditions 
$\mathcal{A}_{2}=0$ and $\mathcal{A}_{5}=0$ are equivalent and 
lead to $\alpha_4=\alpha_5$. Based on this, the requirement $\mathcal{A}_{11}=0$,
leads to: 
\begin{equation}
1+\alpha_4 = \frac{\partial \mathbb{K}[2]}{\bar{k}} = \cos(2\bar{\mu} \gamma\bar{k}) =: \Omega.
\end{equation}
For the sake of simplicity we have defined here the $\Omega$-function.
With use of this, the condition $\mathcal{A}_{6}=0$ leads to 
\begin{equation}
\alpha_6 = \mathbb{K}[2](2+\alpha_9)-\mathbb{K}[s_2]-\bar{k}\Omega. \label{alpha6}
\end{equation}
So, equation (\ref{A12}) simplifies to 
\begin{equation}
\mathcal{A}_{12} =\alpha_9\mathbb{K}[2].  
\end{equation}
Therefore, requiring $\mathcal{A}_{12}=0$ is equivalent to the 
condition $\alpha_9=0$. Furthermore, $\mathcal{A}_{4}=0$ gives $\alpha_7=\alpha_8$. 
The expression for $\alpha_7$ can be derived from the condition $\mathcal{A}_{3}=0$.
Namely, using Eq. (\ref{alpha6}) one obtains: 
\begin{equation}
\alpha_7 = 2(2\beta-1)\mathbb{K}[1]^2+4(1-\beta) \bar{k}\mathbb{K}[2]-2\bar{k}^2\Omega.
\end{equation}
The condition $\mathcal{A}_{13}=0$ is fulfilled by using the expressions  derived
for $\alpha_1$, $\alpha_2$ and $\alpha_6$. The last two anomalies (\ref{A7}) and (\ref{A8}) 
can be simplified to: 
\begin{eqnarray}
\mathcal{A}_7 &=&  2(1+2\beta)(\Omega\mathbb{K}[1]^2-\mathbb{K}[2]^2 ),  \\ 
\mathcal{A}_8 &=&  \bar{k} (1+2\beta)(\mathbb{K}[2]^2-\Omega\mathbb{K}[1]^2).
\end{eqnarray}
The anomaly freedom conditions for those last terms, $\mathcal{A}_{7}=0$ and $\mathcal{A}_8=0$,
are fulfilled if and only if $\beta = -1/2$.  

It is also worth noticing that the function $\mathcal{B}$ given by Eq. (\ref{Bfunction}) 
is equal to zero when the expression obtained for $\alpha_6$ is used. There is 
finally no contribution from the diffeomorphism constraint in the $\left\{H^Q_G, D_G \right\}$ bracket.

Using the anomaly freedom conditions given above, the bracket between the total Hamiltonian 
constraints simplifies to  
\begin{eqnarray}
\left\{ H_{tot}[N_1],H_{tot}[N_2] \right\}  &=& \Omega  D_{tot} \left[ \frac{\bar{N}}{\bar{p}} \partial^a(\delta N_2 - \delta N_1)\right]  \nonumber \\
     &+&   (\alpha_{10}-\alpha_4) D_M \left[ \frac{\bar{N}}{\bar{p}}\partial^a(\delta N_2-\delta N_1)  \right].
\end{eqnarray}
The closure of the algebra of total constraints implies the last condition $\alpha_{10}=\alpha_4=\Omega-1$.\\

To summarize, the counter-terms allowing the algebra to be anomaly-free are uniquely determined, 
and are given by:
\begin{eqnarray}
\alpha_1 &=&   \mathbb{K}[2]-\mathbb{K}[s_1], \label{alpha1}  \\
\alpha_2 &=&   2\mathbb{K}[1]^2 -2\bar{k}\mathbb{K}[2], \\
\alpha_3 &=&  0,  \\
\alpha_4 &=&  \Omega-1,  \\
\alpha_5 &=&  \Omega-1,  \\
\alpha_6 &=&  2\mathbb{K}[2]-\mathbb{K}[s_2]-\bar{k}\Omega,  \label{alpha6}  \\
\alpha_7 &=&  - 4 \mathbb{K}[1]^2+6 \bar{k}\mathbb{K}[2]-2\bar{k}^2\Omega,  \\
\alpha_8 &=&  - 4 \mathbb{K}[1]^2+6 \bar{k}\mathbb{K}[2]-2\bar{k}^2\Omega, \\
\alpha_9 &=&  0,  \\
\alpha_{10} &=&   \Omega-1.
\end{eqnarray}
It is straightforward to check that the counter-terms $\alpha_1,\dots, \alpha_{10}$ are vanishing 
in the classical limit ($\bar{\mu}\rightarrow 0$), as expected. 

Those counter-terms are defined 
up to the two integers $s_1$ and $s_2$, which appear in (\ref{alpha1}) and (\ref{alpha6}). However, 
in the Hamiltonian (\ref{HolModHam}), the factor $\alpha_1$ appears with $\mathbb{K}[s_1]$ 
and the factor $\alpha_6$ appears with $\mathbb{K}[s_2]$. Namely, we have $\mathbb{K}[s_1]+\alpha_1=\mathbb{K}[2]$
and $\mathbb{K}[s_2]+\alpha_6= 2\mathbb{K}[2]-\bar{k}\Omega$. Therefore, the final Hamiltonian 
will not depend on the parameters $s_1$ and $s_2$. No ambiguity remains to be fixed. 

Moreover, the anomaly cancellation requires
\begin{equation}
\beta = -\frac{1}{2},
\end{equation}
which fixes the functional  form of the $\bar{\mu}$ factor. The fact that anomaly freedom requires 
$\beta=-1/2$ is a quite surprising result. The exact value of $\beta$ is highly debated  
in LQC. The only {\it a priori} obvious statement is that $\beta \in [-1/2,0]$. The choice $\beta=-1/2$ is called the
$\bar{\mu}-$scheme (new quantization scheme) and is preferred by some authors for physical reasons 
\cite{corisingh-beta}.  Our result seems to show that the  $\bar{\mu}-$scheme is embedded in the structure 
of the theory and this gives a new motivation for this particular choice of quantization scheme. 
The quantity $\bar{\mu}^2 \bar{p}$ can be interpreted as the physical area of an elementary 
loop along which the holonomy is calculated. Because, in the $\bar{\mu}-$scheme,  
$\bar{\mu}^2  \propto \bar{p}^{-1}$, the physical area of the loop remains constant. 
This elementary area is usually set to  be the area gap $\Delta$ derived in LQG.
Therefore, in the $\bar{\mu}-$scheme,
\begin{equation}
\bar{\mu} = \sqrt{\frac{\Delta}{\bar{p}}}.
\end{equation}

\subsection{Algebra of constraints}

\begin{figure}[htb] 
\includegraphics[scale=0.7]{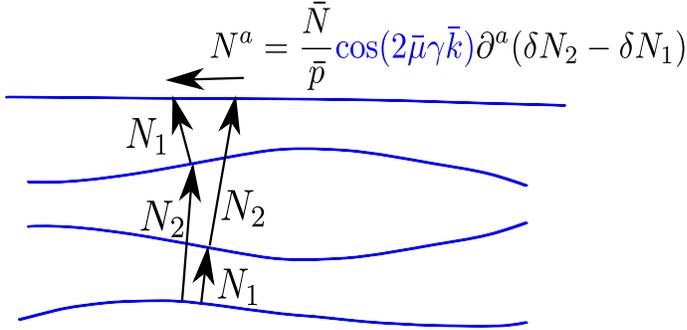} 
\label{FIG1}
\caption{Pictorial representation of the hypersurface deformation algebra (\ref{HtotHtot}).}
\end{figure}

Taking into account the previous conditions of anomaly-freedom, the non-vanishing Poisson brackets 
for the gravity sector are:
\begin{eqnarray}
\left\{ H_G^Q[N],D_G[N^a]\right\} &=& - H_G^Q[\delta N^a \partial_a \delta N],  \\
\left\{ H_G^Q[N_1],H_G^Q[N_2]\right\} &=& \Omega D_G \left[ \frac{\bar{N}}{\bar{p}} \partial^a(\delta N_2 - \delta N_1) \right]. 
\end{eqnarray}
This clearly shows that the \emph{gravity sector is anomaly free}.  The remaining non-vanishing brackets are:
\begin{eqnarray}
\left\{ H_M[N],D_{tot}[N^a] \right\} &=& - H_M [\delta N^a \partial_a \delta N],  \\
\left\{ H_M[N_1],H_M[N_2]\right\} &=& \Omega D_M\left[ \frac{\bar{N}}{\bar{p}} \partial^c(\delta N_2 - \delta N_1)\right]. 
\end{eqnarray}
The algebra of total constraints therefore takes the following form:
\begin{eqnarray}
\left\{ D_{tot}[N^a_1],D_{tot} [N^a_2] \right\} &=& 0, \\
\left\{ H_{tot}[N],D_{tot}[N^a] \right\} &=& - H_{tot}[\delta N^a \partial_a \delta N], \\
\left\{ H_{tot}[N_1],H_{tot}[N_2] \right\} &=&  D_{tot} \left[  \Omega  \frac{\bar{N}}{\bar{p}} \partial^a(\delta N_2 - \delta N_1)\right].  
\label{HtotHtot}
\end{eqnarray}
Although the algebra is closed, there are however modifications with respect to the classical case, 
due to presence of the factor $\Omega$ in Eq. (\ref{HtotHtot}). Therefore, not only 
the dynamics, as a result of the modification of the Hamiltonian constraint, is modified but also the very structure of the space-time itself is \emph{deformed}. This is embedded 
in the form of the algebra of constraints. The hypersurface  deformation algebra generated 
by (\ref{HtotHtot}) is pictorially represented in Fig. \ref{FIG1}.
As $\Omega \in [-1,1]$, the shift vector 
\begin{equation}
N^a=  \Omega  \frac{\bar{N}}{\bar{p}} \partial^a(\delta N_2 - \delta N_1)
\label{ShiftvectOmega} 
\end{equation}
appearing in (\ref{HtotHtot}) can change sign in time.

 In order to see when this might
happen let us express the parameter $\Omega$ as: 
\begin{equation}
\Omega = \cos(2\bar{\mu} \gamma\bar{k}) = 1 - 2\frac{\rho}{\rho_c},
\end{equation}
where $\rho$ is the energy density of the matter field and 
\begin{equation}
\rho_c = \frac{3}{\kappa \gamma \bar{\mu}^2 \bar{p}}  = \frac{3}{\kappa \gamma \Delta}.
\end{equation}
In the low energy limit, $\rho \rightarrow 0$, the classical case $(\Omega \rightarrow 1)$ is 
correctly recovered. However, while approaching the high energy domain the situation 
drastically changes. Namely, for $\rho = \rho_c/2$, the  shift vector (\ref{ShiftvectOmega}) 
becomes null. At this point, the maximum value of the Hubble parameter is also reached.  
The maximum allowed energy density is $\rho = \rho_c$ and corresponds to the
bounce. Then the shift vector (\ref{ShiftvectOmega})  fully reverses with respect to 
the low energy limit.  One can interpret this peculiar behavior as a geometry
change. Namely, when the universe is in its quantum stage ($\rho > \rho_c/2$), the effective 
algebra of constraints shows that the space is Euclidian.  At the particular value 
$\rho = \frac{\rho_c}{2}$, the geometry switches to the Minkowski one \cite{private_discussion}. 
This will become even clearer when analyzing the Mukhanov equation in Sec. \ref{equations}.
The consequences of this have not yet been fully understood, but it is interesting to notice that this model 
naturally exhibits properties related to the Hartle-Hawking no-boundary proposal \cite{Hartle:1983ai}. 

\section{Equations of motion} \label{equations}

Once the anomaly-free theory of scalar perturbations with holonomy corrections is constructed, 
the equations of motion for the canonical variables can be derived. This can be achieved through 
the Hamilton equation
\begin{equation}
\dot{f} = \{f ,H[N,N^a]\}  \label{HamEq},
\end{equation}
where the Hamiltonian $H[N,N^a]$ is the sum of all constraints 
\begin{equation}
H[N,N^a] =  H^Q_{G}[N]+H_{M}[N]+D_{G} [N^a]+D_{M} [N^a].  
\end{equation}

\subsection{Background equations}

Based on the Hamilton equation (\ref{HamEq}), the equations for the 
canonical background variables are the following:
\begin{eqnarray}
\dot{\bar{k}} &=& -\frac{\bar{N}}{2\sqrt{\bar{p}}}\mathbb{K}[1]^2-\bar{N}\sqrt{\bar{p}} \frac{\partial}{\partial \bar{p}} \mathbb{K}[1]^2 
+\frac{\kappa}{2} \sqrt{\bar{p}} \bar{N}\left[  -\frac{\bar{\pi}^2}{2\bar{p}^3}+V(\bar{\varphi}) \right], \label{dotkbar}    \\
\dot{\bar{p}} &=&  2 \bar{N} \sqrt{\bar{p}}\mathbb{K}[2], \label{dotpbar} \\
\dot{\bar{\varphi}} &=&   \bar{N} \frac{\bar{\pi}}{\bar{p}^{3/2}},  \label{dotbarvarphi} \\
\dot{\bar{\pi}} &=&  -\bar{N}  \bar{p}^{3/2} V_{, \varphi}(\bar{\varphi}). \label{dotbarpi}
\end{eqnarray}
In the following, we choose the time to be conformal by setting $\bar{N}=\sqrt{\bar{p}}$. 
The $``\cdot"$ then means differentiation with respect to conformal time $\eta$.
 
Eqs. (\ref{dotbarvarphi}) and (\ref{dotbarpi}) can be now combined into the Klein-Gordon equation
\begin{equation}
\ddot{\bar{\varphi}}+2 \mathbb{K}[2] \dot{\bar{\varphi}}+\bar{p} V_{,\varphi} (\bar{\varphi}) = 0.
\end{equation}
Eq. (\ref{dotpbar}), together with the background part of the Hamiltonian constraint
\begin{equation}
\frac{1}{V_0} \frac{\partial H}{\partial \bar{N}} = \frac{1}{2\kappa}\left[-6 \sqrt{\bar{p}} (\mathbb{K}[1])^2\right]
+ \bar{p}^{3/2}\left[\frac{\bar{\pi}^2}{2\bar{p}^{3}}+V(\bar{\varphi})\right]= 0, \label{HamConstBack}
\end{equation}
lead to the modified Friedmann equation 
\begin{equation}
\mathcal{H}^2 =\bar{p} \frac{\kappa}{3} \rho \left(1-\frac{\rho}{\rho_c}\right).
\end{equation}
Another useful expression is:
\begin{equation}
3 \mathbb{K}[1]^2 =  \frac{\bar{\pi}^2}{2\bar{p}^2} + \bar{p} V(\bar{\varphi}).
\end{equation}
Here $\mathcal{H}$ stands for the conformal Hubble factor 
\begin{equation}
\mathcal{H} := \frac{\dot{\bar{p}}}{2\bar{p}} = \mathbb{K}[2].
\end{equation}
The energy density and pressure of the scalar field are given by: 
\begin{eqnarray}
\rho &=& \frac{\bar{\pi}^2}{2\bar{p}^{3}}+V(\varphi),   \\
P &=& \frac{\bar{\pi}^2}{2\bar{p}^{3}}-V(\varphi).
\end{eqnarray}  
For the purpose of further considerations, we also derive the relation
\begin{equation}
\kappa \left( \frac{\bar{\pi}^2}{2\bar{p}^2}\right) = \bar{k} \mathbb{K}[2]-\dot{\bar{k}},
\end{equation}
which comes from Eq. (\ref{dotkbar}) combined with (\ref{HamConstBack}).  

\subsection{Equations for the perturbed variables}

The equations for the perturbed parts of the canonical variables are:
\begin{eqnarray}
\delta \dot{E}^a_i&=& -\bar{N}\left[ 
\sqrt{\bar{p}} \Omega \delta K^j_c \delta^c_i \delta^a_j- \sqrt{\bar{p}} \Omega  (\delta K^j_c \delta^c_j) \delta^a_i 
- \frac{1}{\sqrt{\bar{p}}} (2\mathbb{K}[2]-\bar{k}\Omega)  \delta E^a_i \right] +\nonumber \\
&+&\delta N\left(2  \mathbb{K}[2]  \sqrt{\bar{p}}  \delta^a_i \right) 
-\bar{p} (\partial_i \delta N^a -(\partial_c \delta N^c) \delta^a_i),  \label{dotdE}  \\
\delta \dot{K}^i_a &=& \bar{N} \left[- \frac{1}{\sqrt{\bar{p}}} (2\mathbb{K}[2]-\bar{k}\Omega) \delta K^i_a \right.  \nonumber \\
&-&\frac{1}{2 \bar{p}^{3/2}}(- 3 \mathbb{K}[1]^2+6 \bar{k}\mathbb{K}[2]-2\bar{k}^2\Omega)\delta E^c_j \delta ^j_a \delta^i_c   \nonumber \\
&+& \left. \frac{1}{4 \bar{p}^\frac{3}{2}}(- 3 \mathbb{K}[1]^2+6 \bar{k}\mathbb{K}[2]-2\bar{k}^2\Omega)(\delta E^c_j \delta^j_c) \delta^i_a
+\frac{\delta^{ik}}{2 \bar{p}^\frac{3}{2}} \partial_a \partial_d \delta E^d_k\right] \nonumber \\
&+& \frac{1}{2} \left[ -\frac{1}{\sqrt{\bar{p}}} (3\mathbb{K}[1]^2-2\bar{k}\mathbb{K}[2]) \delta^i_a  \delta N
+\frac{2}{\sqrt{\bar{p}}} (\partial_a \partial^i \delta N) \right]  \nonumber \\
&+& \delta^i_c (\partial_a \delta N^c)+ \kappa \delta N \frac{\sqrt{\bar{p}}}{2} \left[ - \frac{\bar{\pi}^2}{2\bar{p}^{3}}+V(\bar{\varphi}) \right] \delta^i_a     \nonumber \\
 &+& \kappa \bar{N} \left[  -\frac{\bar{\pi} \delta \pi}{2\bar{p}^{5/2}} \delta^i_a
 +\frac{\sqrt{\bar{p}}}{2} \delta \varphi  \frac{\partial V (\bar{\varphi}) }{\partial \bar{\varphi}} \delta^i_a  
 +  \left(  \frac{\bar{\pi}^2}{2\bar{p}^{3/2}} + \bar{p}^{3/2} V(\bar{\varphi}) \right)\frac{\delta^j_c \delta E^c_j}{4\bar{p}^2}\delta^i_a \right.   \nonumber \\
 &+& \left. \left(  \frac{\bar{\pi}^2}{2\bar{p}^{3/2}} - \bar{p}^{3/2} V(\bar{\varphi}) \right)\frac{\delta^i_c  \delta^j_a \delta E^c_j}{2\bar{p}^2}  \right],  
  \label{dotdK}  \\
\delta \dot{\varphi} &=&  \delta N \left(  \frac{\bar{\pi}}{ \bar{p}^{3/2}}\right) 
+\bar{N} \left(\frac{\delta \pi}{\bar{p}^{3/2}} -\frac{\bar{\pi}}{\bar{p}^{3/2}} \frac{\delta^j_c \delta E^c_j}{2\bar{p}} \right), \label{dotdeltavarphi} \\
\delta \dot{\pi} &=& -\delta N \left(\bar{p}^{3/2} V_{,\varphi}(\bar{\varphi})  \right)+\bar{\pi}(\partial_a \delta N^a) \nonumber \\
&-&\bar{N}\left[ - \sqrt{\bar{p}} \Omega \delta^{ab} \partial_a\partial_b\delta \varphi+\bar{p}^{3/2} V_{,\varphi\varphi}(\bar{\varphi}) \delta \varphi 
+\bar{p}^{3/2} V_{,\varphi}(\bar{\varphi}) \frac{\delta^j_c \delta E^c_j}{2\bar{p}}   \right]. \label{dotdeltapi}
\end{eqnarray}

\subsection{Longitudinal gauge}

As an example of application we will now derive the equations in the longitudinal gauge.  
In this case, the $E$ and $B$ perturbations are set to  zero. 
The line element  (\ref{lineelement}) therefore simplifies to 
\begin{equation}
ds^2 = a^2\left[ -(1+2\phi)d\eta^2+(1-2\psi)\delta_{ab}dx^adx^b  \right],
\end{equation}
where $\phi$ and $\psi$ are two remaining perturbation functions and  $a$ is the 
scale factor. From the metric above, one can 
derive the laps function, the shift vector and the spatial metric: 
\begin{eqnarray}
N &=& a\sqrt{1+2\phi},  \\
N^a &=& 0,  \\
q_{ab} &=& a^2(1-2\psi)\delta_{ab}. \label{qab}
\end{eqnarray}
The lapse function can be expanded for the background 
and perturbation part as $N=\bar{N} +\delta N$, where 
\begin{eqnarray}
\bar{N} &=& \sqrt{\bar{p}} = a,  \\
\delta N &=& \bar{N} \phi.
\end{eqnarray}
Using Eq. (\ref{qab}), the perturbation of the densitized triad is expressed as:
\begin{equation}
\delta E^a_i=-2\bar{p} \psi \delta^a_i.  \label{dEpsi}
\end{equation}
The time derivative of this expression will also be useful and can be written as:  
\begin{equation}
\delta \dot{E}^a_i = -2\bar{p} (2\mathbb{K}[2]\psi+\dot{\psi})\delta^a_i. 
\end{equation}
Let us now find the expression for the perturbation of the extrinsic curvature $\delta K^i_a$
in terms of the metric perturbations $\phi$ and $\psi$. For this purpose, one can apply the
expression  (\ref{dEpsi}) to the left hand side of (\ref{dotdE}). The resulting equation can be 
solved for $\delta K^i_a$, leading to:
\begin{equation}
\delta K^i_a = - \delta^i_a \frac{1}{\Omega} \left(  \dot{\psi}+\bar{k}\Omega \psi +\mathbb{K}[2]\phi \right).
\end{equation}
The time derivative of this variable is given by
\begin{eqnarray}
\delta \dot{K}^i_a = \delta^i_a \frac{1}{\Omega} \left[ -\ddot{\psi}  -\dot{\bar{k}}\Omega \psi 
+\dot{\psi}\left(  \frac{\dot{\Omega}}{\Omega}-\bar{k}\Omega \right)
+\phi \mathbb{K}[2] \frac{\dot{\Omega}}{\Omega} \right.  \nonumber \\
- \left. \phi \dot{\mathbb{K}}[2]-\mathbb{K}[2]\dot{\phi} \right]. \label{ddK}
\end{eqnarray}
Applying (\ref{ddK}) to the left hand side of (\ref{dotdK}), the equation containing the diagonal 
part as well as the off-diagonal contribution is easily obtained. The off-diagonal part leads to 
\begin{equation}
\partial_a\partial^i(\phi-\psi) = 0.   
\end{equation} 
This translates into $\psi=\phi$.  In what follows, we will therefore consider $\phi$ 
only. The diagonal part of the discussed equation can be expressed as:
\begin{eqnarray}
 \ddot{\phi}&+&  \dot{\phi}\left[ 3\mathbb{K}[2]- \frac{\dot{\Omega}}{\Omega}\right]
+ \phi \left[ \dot{\mathbb{K}}[2] +2\mathbb{K}[2]^2  -\mathbb{K}[2] \frac{\dot{\Omega}}{\Omega} \right]  \nonumber \\
&=& 4\pi G \Omega \left[ \dot{\bar{\varphi}} \delta \dot{\varphi}-\bar{p} \delta \varphi  V_{,\varphi} (\bar{\varphi})\right].
\label{eqfin1}
\end{eqnarray}
One case now use the diffeomorphism constraint
\begin{equation}
\kappa \frac{\delta H[N,N^a]}{\delta (\delta N^c)} = \bar{p} \partial_c(\delta^d_k \delta K^k_d )
-\bar{p}(\partial_k \delta K^k_c)
-\bar{k}\delta^k_c (\partial_d \delta E^d_k  )
+\kappa \bar{\pi} (\partial_c \delta \varphi) = 0. 
\end{equation}
With the expressions for $\delta K^i_a$ and  $\delta E^a_i$, it can be derived that
\begin{equation}
\partial_c \left[\dot{\phi}+\phi \mathbb{K}[2] \right] = 4\pi G \Omega  \dot{\bar{\varphi}} \partial_c \delta \varphi.
\label{eqfin2}
\end{equation}
The next equation comes from the perturbed part of the Hamiltonian constraint:
\begin{eqnarray}
\frac{\delta H[N,N^a]}{\delta (\delta N)} &=& 
\frac{1}{2\kappa} \left[-4\sqrt{\bar{p}} \mathbb{K}[2] \delta^c_j \delta K^j_c
-\frac{1}{\sqrt{\bar{p}}}  \left( 3\mathbb{K}[1]^2-2\bar{k}\mathbb{K}[2]  \right) \delta^j_c\delta E^c_j  \right. \nonumber \\
&+&\left. \frac{2}{\sqrt{\bar{p}}} \partial_c \partial^j \delta E^c_j \right]  
+ \frac{\bar{\pi} \delta \pi}{\bar{p}^{3/2}}-\frac{\bar{\pi}^2}{2\bar{p}^{3/2}} \frac{\delta^j_c \delta E^c_j}{2\bar{p}} \nonumber \\
&+&\bar{p}^{3/2} \left[ V_{,\varphi}(\bar{\varphi}) \delta \varphi +V(\bar{\varphi}) \frac{\delta^j_c \delta E^c_j}{2\bar{p}} \right] = 0.
\end{eqnarray}
Using the expressions for $\delta K^i_a$ and  $\delta E^a_i$, this can be rewritten as:
\begin{eqnarray}
\Omega \nabla^2 \phi -3\mathbb{K}[2] \dot{\phi} - \left[  \dot{\mathbb{K}}[2]+2\mathbb{K}[2]^2 \right] \phi
= 4\pi G \Omega \left[ \dot{\bar{\varphi}} \delta \dot{\varphi}+\bar{p} \delta \varphi  V_{,\varphi} (\bar{\varphi})\right].
\label{eqfin3}
\end{eqnarray}
The last equality comes from  (\ref{dotdeltavarphi}) and  (\ref{dotdeltapi}):
\begin{equation}
 \delta \ddot{\varphi}+2\mathbb{K}[2]  \delta \dot{\varphi}-\Omega \nabla^2 \delta \varphi 
+\bar{p}V_{,\varphi\varphi}(\bar{\varphi} ) \delta \varphi
+2 \bar{p} V_{,\varphi}(\bar{\varphi} ) \phi -4\dot{\bar{\varphi}} \dot{\phi} = 0.
\label{eqfin4}
\end{equation}
 
The equations (\ref{eqfin1}), (\ref{eqfin2}) and (\ref{eqfin3})  can be now combined into:
\begin{eqnarray}
\ddot{\phi}+2\left[\mathcal{H}- \left(\frac{\ddot{\bar{\varphi}}}{\dot{\bar{\varphi}}}+\epsilon \right)\right]\dot{\phi} 
+2\left[ \dot{\mathcal{H}}-\mathcal{H} \left( \frac{\ddot{\bar{\varphi}}}{\dot{\bar{\varphi}}}+\epsilon\right) \right]\phi-c_s^2\nabla^2\phi=0,  
\label{finaleq}
\end{eqnarray}
with the quantum correction
\begin{equation}
\epsilon =  \frac{1}{2} \frac{\dot{\Omega}}{\Omega} =  3 \mathbb{K}[2] \left(\frac{\rho+P}{\rho_c-2\rho} \right),
\end{equation}
and the squared velocity
\begin{equation}
c_s^2 = \Omega. 
\end{equation}
The squared velocity of the  perturbation field $\phi$ is equal to $\Omega$.  Because $-1 \leq \Omega \leq 1$,
the speed of perturbations is never super-luminal. However, for $\Omega < 0$ perturbations become unstable 
($c_s^2 < 0$). This corresponds to the energy density regime $\rho > \frac{\rho_{c}}{2}$, where the phase of 
super-inflation is expected.   

At the point $\rho=\frac{\rho_{c}}{2}$, the velocity of the  perturbation field $\phi$ is vanishing. Therefore, 
perturbations don't propagate anymore when approaching  $\rho=\frac{\rho_{c}}{2}$, where the Hubble factor 
reaches its maximal value. Moreover, at this point, the quantum correction $\epsilon \rightarrow \infty$. 
Because of this, Eq.  (\ref{finaleq}) diverges and cannot be used to determine the propagation 
of the perturbations. However, as shown in the next section, the equation for the gauge-invariant Mukhanov  variable
does not exhibit such a pathology.   

It is interesting to notice that the equations of motion derived in this subsection are the same as those 
found in \cite{WilsonEwing:2011es}. This is quite surprising, because they were derived 
following independent paths. In our approach, we have introduced the most general form for the holonomy corrections 
to the Hamiltonian. Then, by adding counter-terms, anomalies in the algebra of constraints 
were removed. On the other hand, the method proposed in  \cite{WilsonEwing:2011es} is based on the diagonal form of the metric in the 
longitudinal gauge. This enables one to introduce holonomy corrections in almost  the same way as  in the case 
of a homogeneous model. It was then shown that a system defined in this way stays on-shell, that is, is free 
of anomalies. The non-trivial equivalence of both approaches may suggest uniqueness in defining a
theory of scalar perturbations with holonomy corrections in an anomaly-free manner.
   
\subsection{Gauge invariant variables and Mukhanov equation} \label{GIVaMe}

Considering the scalar perturbations, there is only one physical degree of freedom. 
As it was shown in \cite{Mukhanov:1990me}, this physical variable combines 
both the perturbation of the metric and the perturbation of matter.  The classical expression 
on this gauge-invariant quantity is:  
\begin{equation}
v = a(\eta) \left( \delta \varphi^{GI} + \frac{\dot{\bar{\varphi}}}{\mathcal{H}} \Psi \right),
\label{MukhClass}
\end{equation}
and its equation of motion is given by
\begin{equation}
\ddot{v}-\nabla^2 v - \frac{\ddot{z}}{z} v = 0,
\end{equation}
where 
\begin{equation}
z = a(\eta) \frac{\dot{\bar{\varphi}}}{\mathcal{H}}.
\end{equation}

In the canonical formalism with scalar perturbations, the gauge transformation of a variable 
$X$ under a small coordinate transformation
\begin{equation}
x^{\mu} \rightarrow  x^{\mu} + \xi^{\mu} \ \ \ \ ; \ \ \ \ \xi^{\mu} = (\xi^0, \partial^a \xi),
\end{equation}
is given by (see \cite{Bojowald:2008jv}  for details):
\begin{equation}
\delta_{[\xi^0, \xi]} X \dot{=} \{X, H^{(2)}[\bar{N}\xi^0] + D^{(2)}[\partial^a \xi] \},
\end{equation}
and it is straightforward to see that, classically,
\begin{equation}
\delta_{[\xi^0, \xi]} v = 0.
\end{equation}
This means that $v$ is diffeomorphism-invariant and can be taken as an observable. 

Taking into account the holonomy corrections introduced in this paper, 
the $\Omega$ function will  modify the gauge  transformations of the time 
derivative of a variable $X$, so that
\begin{equation}
\delta_{[\xi^0, \xi]} \dot{X} - (\delta_{[\xi^0, \xi]} X) \dot{} = \Omega \cdot \delta_{[0, \xi^0]} X.
\end{equation}
Using this relation and gauge transformations of the metric perturbations
\begin{eqnarray} 
\delta_{[\xi^0, \xi]}\psi &=& -\mathbb{K}[2] \xi^0, \\
\delta_{[\xi^0, \xi]}\phi  &=& \dot{\xi}^0+\mathbb{K}[2]\xi^0,  \\
\delta_{[\xi^0, \xi]}E &=& \xi, \\
\delta_{[\xi^0, \xi]}B  &=& \dot{\xi}, 
\end{eqnarray}
one can define the gauge-invariant variables (Bardeen potentials) as:
 \begin{eqnarray}
\Phi &=& \phi +\frac{1}{\Omega} (\dot{B}-\ddot{E}) +\left( \frac{\mathbb{K}[2]}{\Omega}-\frac{\dot{\Omega}}{\Omega}\right)  (B-\dot{E}),  \\
\Psi &=& \psi - \frac{\mathbb{K}[2]}{\Omega} (B-\dot{E}), \\
\delta \varphi^{GI} &=& \delta \varphi + \frac{\dot{\bar{\varphi}}}{\Omega} (B-\dot{E}).
\end{eqnarray}
The normalization of these variables  was set such that, in the longitudinal gauge $(B=0=E)$, 
we have $\Phi = \phi $, $\Psi = \psi $ and $\delta \varphi^{GI} = \delta \varphi $.
It is possible to define the analogous of the Mukhanov variable (\ref{MukhClass}): 
\begin{equation}
v := \sqrt{\bar{p}} \left( \delta \varphi^{GI} + \frac{\dot{\bar{\varphi}}}{\mathbb{K}[2]} \Psi \right).
\label{MukhQuant}
\end{equation}
Writing the equations for $\Psi$ and $\delta \varphi^{GI}$, which are 
\begin{eqnarray}
\ddot{\Psi}+2\left[\mathcal{H}- \left(\frac{\ddot{\bar{\varphi}}}{\dot{\bar{\varphi}}}+\epsilon \right)\right]\dot{\Psi} 
+2\left[ \dot{\mathcal{H}}-\mathcal{H} \left( \frac{\ddot{\bar{\varphi}}}{\dot{\bar{\varphi}}}+\epsilon\right) \right]\Psi-c_s^2\nabla^2\Psi=0  
\end{eqnarray}
and
\begin{equation}
 \delta \ddot{\varphi}^{GI}+2\mathbb{K}[2]  \delta \dot{\varphi}^{GI}-\Omega \nabla^2 \delta \varphi^{GI}
+\bar{p}V_{,\varphi\varphi}(\bar{\varphi} ) \delta \varphi^{GI}
+2 \bar{p} V_{,\varphi}(\bar{\varphi} ) \Psi -4\dot{\bar{\varphi}}^{GI} \dot{\Psi} = 0,
\end{equation}
one obtains equation for the variable (\ref{MukhQuant}):
\begin{eqnarray}
&& \ddot{v}-\Omega \nabla^2 v - \frac{\ddot{z}}{z} v = 0, \label{HCMukhequation}\\
&& z = \sqrt{\bar{p}} \frac{\dot{\bar{\varphi}}}{\mathbb{K}[2]},
\end{eqnarray}
which corresponds to the Mukhanov equation for our model. As we see, the difference between the 
classical and the holonomy-corrected case is the factor $\Omega$ in front of the Laplacian. 
This quantum contribution leads to a variation of the propagation velocity of the perturbation $v$. This is
similar to the case of the perturbation $\phi$ considered in the previous subsection. 
The main difference is that there is no divergence for $\rho = \rho_c/2$ and the evolution of perturbations
can be investigated in the regime of high energy densities. It is once again worth noticing that for $\rho > \rho_c/2$,
$\Omega$ becomes negative and Eq. (\ref{HCMukhequation}) changes  from an
hyperbolic form to an elliptic one. This basically means that the time part becomes indistinguishable
from the spatial one. This can be interpreted as a transition from a Minkowskian geometry to and Euclidean 
geometry, as mentioned earlier.

Finally, it is also possible to define the perturbation of curvature $\mathcal{R}$ such that
\begin{equation}
\mathcal{R} = \frac{v}{z}.
\end{equation}
Based on this,  one can now calculate the power spectrum of scalar perturbations. This 
opens new possible ways to study quantum gravity effects in the very early universe. 
Promising applications of the derived equations 
will be investigated elsewhere. 

\section{Summary and conclusions} \label{summary}

In this paper, we have investigated the theory of scalar perturbations with
holonomy corrections. Such corrections are expected because of quantum 
gravity effects predicted by LQG. They basically come from the
regularization of the curvature of the connection at the Planck scale. Because
of this, the holonomy corrections become dominant in the high curvature regime.
The introduction of "generic type" holonomy corrections leads to an anomalous 
algebra of constraints.  The conditions of anomaly freedom impose some 
restrictions on the form of the holonomy corrections. However, we have shown
that the holonomy corrections, in the standard form, cannot fully satisfy the 
conditions of anomaly freedom. In order to solve this difficulty, additional counter-terms
were introduced. Such counter-terms tend to zero in the classical limit, 
but play the role of regularizators of anomalies in the quantum (high curvature) 
regime. The method of  counter-terms was earlier successfully applied to
cosmological perturbations with inverse-triad corrections \cite{Bojowald:2008gz}.
 
We have shown that, thanks to the  counter-terms, the theory of cosmological 
perturbations with holonomy corrections can be formulated in an  anomaly-free
way. The anomaly freedom was shown to be fulfilled not only for the gravity 
sector but also when taking into account scalar matter. The requirements of 
anomaly freedom were used to determine the form of the counter-terms. Furthermore, 
conditions for obtaining and anomaly-free algebra of constraints were shown to be fulfilled 
only for a particular choice of the $\bar{\mu}$ function, namely for the 
$\bar{\mu}-$scheme (new quantization scheme). This quantization scheme was 
shown earlier to be favored because of the consistency of the 
background dynamics \cite{corisingh-beta}.  Our result supports these earlier claims. 

In our formulation, the diffeomorphism constraint holds its classical form,
in agreement with the LQG expectations. The obtained anomaly-free gravitational 
Hamiltonian contains seven holonomy modifications. It was also necessary to 
introduce one counter-term into the matter Hamiltonian in order to ensure the closure 
of the algebra of total constraints. There is no ambiguity in defining the holonomy 
corrections after imposing the anomaly-free conditions. The only remaining free 
parameter of the theory is the area gap $\Delta$ used in defining the $\bar{\mu}$ 
function. This quantity can however be possibly fixed with the spectrum 
of the area operator in LQG. Based on the equations derived in this paper it will  
also be possible to put observational constraints on the value of $\Delta$ and, 
hence, on the critical energy density $\rho_c$.

Based on the studied anomaly-free formulation, equations of motion were derived.
As an example of application, we studied the equations in the longitudinal gauge.
We have also found the gauge-invariant variables, which are holonomy-corrected
versions of the Bardeen potentials. Using this, we have derived the equation for the 
Mukhanov variable.   This equation can be directly used to compute the power 
spectrum of scalar perturbations with quantum gravitational holonomy corrections.
Similar considerations were studied in the case of inverse-triad corrections 
\cite{arXiv:1011.2779}. In that case,  observational consequences have been derived 
and compared with CMB data \cite{Bojowald:2011hd, arXiv:1107.1540}.
  
\ack

The authors would like to thank M.~Bojowald, G.~Calcagni and E.~Wilson-Ewing 
for the discussions. TC and JM were supported from the Astrophysics Poland-France 
(Astro-PF). JM has been supported by Polish Ministry of Science and Higher Education 
grant N N203 386437 and by Foundation of Polish Science award START.

\end{document}